\documentclass[12pt]{article}

\usepackage{amsmath}
\usepackage{amssymb}
\usepackage{latexsym}
\usepackage{graphicx}
\usepackage{psfrag,fancyhdr,epsfig}

\def\beq{\begin{equation}}
\def\eeq{\end{equation}}
\def\bea{\begin{eqnarray}}
\def\eea{\end{eqnarray}}

\begin{document}

\begin{titlepage}

\hfill hep-th/0601126

\vspace*{1.5cm}
\begin{center}
{\bf \Large Graviton Emission in the Bulk from a\\[4mm] Higher-Dimensional
Schwarzschild Black Hole}

\bigskip \bigskip \medskip

{\bf S. Creek}$^1$, {\bf O. Efthimiou}$^2$, {\bf P. Kanti}$^1$ and
{\bf K. Tamvakis}$^2$

\bigskip
$^1$ {\it Department of Mathematical Sciences, University of Durham,\\
Science Site, South Road, Durham DH1 3LE, United Kingdom}

$^2$ {\it Division of Theoretical Physics, Department of Physics,\\
University of Ioannina, Ioannina GR-45110, Greece}

\bigskip \medskip
{\bf Abstract}
\end{center}
We consider the evaporation of $(4+n)$-dimensional non-rotating black holes
into gravitons. We calculate the energy emission rate for gravitons in the bulk
obtaining analytical solutions of the master equation satisfied by all three types
(S,V,T) of gravitational perturbations. Our results, valid in the low-energy
regime, show a vector radiation dominance for every value of $n$, while the
relative magnitude of the energy emission rate of the subdominant scalar and
tensor radiation depends on $n$. The low-energy emission rate in the bulk
for gravitons is well below that for a scalar field, due to the absence of
the dominant $\ell=0,1$ modes from the gravitational spectrum. Higher partial
waves though may modify this behaviour at higher energies. The calculated
low-energy emission rate, for all types of degrees of freedom decreases with
$n$, although the full energy emission rate, integrated over all frequencies,
is expected to increase with $n$, as in the previously studied case of a
bulk scalar field.

\end{titlepage}

\section{Introduction}

The quest for unification of fundamental interactions with gravity has led
to the idea of extra spatial dimensions. String Theory, or some form of it,
considered at present as a promising consistent theoretical framework for
such a unification at the quantum level, requires extra spatial dimensions.
In this higher-dimensional context, which has received a lot of attention
in the last few years \cite{ADD, RS}, gravity propagates in $D=4+n$
dimensions ({\textit{Bulk}}), while all Standard Model degrees of freedom
are assumed to be confined on a four-dimensional $D$-{\textit{Brane}}. In
theories with large extra dimensions \cite{ADD} the traditional Planck
scale $M_{Pl}\sim 10^{18}$\,GeV is only an effective scale, related to the
fundamental higher-dimensional gravity scale $M_{*}$ through the relation
$M_{Pl}^2 \sim M_{*}^{n+2}R^n$, where $R\sim (V_n)^{1/n}$ is the effective
size of the $n$ extra spatial dimensions. If 
$R \gg \ell_{Pl}\approx 10^{-33}\,cm$, the scale $M_{*}$ can be substantially
lower than $M_{Pl}$. If $M_{*}$ is sufficiently low, trans-planckian particle
collisions could be feasible in accelerators or other environments. The
product of such collisions would be objects that probe the extra dimensions.
Black holes are among the possible products of trans-planckian collisions
\cite{creation}. Black holes with a horizon size $r_H$ smaller than the size
of extra dimensions $R$ will be higher-dimensional objects
centered at the brane and extending in the bulk. A classical treatment
would require that the mass of the black hole $M_{BH}$ would have to be
larger than the fundamental gravity scale $M_{*}$. Such black holes can
be created in ground based colliders, although their appearance in cosmic
rays is possible as well (for references, see the reviews \cite{Kanti, reviews}).
A black hole created in such trans-planckian collisions
is expected to undergo a short {\textit{``balding"}} phase in which it
will shed its additional quantum numbers, apart from charge, angular momentum
and mass. Then, after a more familiar Kerr-like phase, in which it will
lose its angular momentum, a Schwarzschild phase follows during which the
black hole will gradually lose its mass through the emission of Hawking
radiation both in the bulk and in the brane, consisting of elementary
particles of a very distinct thermal spectrum. This black hole radiation
has been the subject of both analytical and numerical studies. This includes
lower spin degrees of freedom \cite{kmr1, FS, HK1, BGK, Barrau, Jung} as
well as graviton emission in the bulk \cite{Naylor}. The study of graviton
emission in the bulk corresponds to the study of perturbations in a
gravitational background. The formalism for the treatment of 
gravitational perturbations of a higher-dimensional non-rotating black
hole has been developed in \cite{IK}. In the present article we consider the evaporation of $(4+n)$-dimensional non-rotating black holes into gravitons
and calculate the energy emission rate for gravitons in the bulk obtaining
analytical solutions of the master equation satisfied by gravitational
perturbations. Our results, valid in the low-energy regime, are complementary
to existing studies in the intermediate energy regime \cite{Naylor}. These
results show a vector radiation dominance for every value of $n$, while the
relative magnitude of the energy emission rate of the subdominant scalar and
tensor radiation depends on $n$. The low-energy emission rate in the bulk
for gravitons is well below that for a scalar field, due to the absence of
the dominant $\ell=0,1$ modes from the gravitational spectrum, although higher
partial waves are likely to modify this behaviour at higher energies. The
calculated low-energy emission rate, for all types of degrees of freedom
decreases with $n$, although the full energy emission rate, integrated over
all frequencies, is expected to increase with $n$, as in the previously
studied case of a bulk scalar field.


\section{Theoretical Framework}

According to the Large Extra Dimensions scenario \cite{ADD}, our universe is made up
of the usual 3+1 non-compact dimensions plus $n$ additional, compact, spacelike ones.
The gravitational background around a spherically-symmetric, neutral
black hole formed in such a $(4+n)$-dimensional, flat spacetime is described by
the following line-element \cite{TMP}
\beq
ds^2 = - f(r)\,dt^2 + \frac{dr^2}{f(r)} + r^2 d\Omega_{2+n}^2,
\label{bhmetric}
\eeq
with the metric function $f(r)$ given by 
\beq
f(r)=1-\left(\frac{r_H}{r}\right)^{n+1}\,, \label{f-fun}
\eeq
and $d\Omega_{2+n}^2$ denoting the area of the ($2+n$)-dimensional unit sphere.
The above black hole is characterized by a non-vanishing temperature given by
\beq
T_H=\frac{n+1}{4 \pi r_H}\,,
\label{TH}
\eeq
and therefore emits Hawking radiation \cite{hawking} with an energy emission
rate that resembles the one for a black body. The strong gravitational field
surrounding the black hole modifies the spectrum through the modification of
the corresponding absorption probability $|{\cal A}|^2$, that now depends on
the energy $\omega$ and spin $s$ of the emitted particle as well as on the
number of extra dimensions $n$ (see, for example, \cite{Kanti}). The absorption
probability can be found by solving the equation of motion of a particular
field propagating in the vicinity of the black hole and by
using classical scattering theory. As mentioned in the Introduction, in this
work, we focus on the decay of a $(4+n)$-dimensional, Schwarzschild
black hole through the emission of gravitons in the bulk. 

Following \cite{IK}, a graviton in $D$ dimensions can be decomposed into
a symmetric traceless tensor, a vector and a scalar part. These are further
expanded in terms of the spin-weighted spherical harmonics on the $S^{n+2}$
unit sphere.
The radial parts of all three types of
gravitational perturbations are found to satisfy a Schrodinger-like equation
of the form \cite{IK}
\beq
f \frac{d}{dr}\left(f\, \frac{d\Phi}{dr}\right) + (\omega^2 -V)\,\Phi=0\,.
\label{tensor}
\eeq
The potential $V$ has a different form for each type of perturbation, namely
\beq
V_{T,V} =  \frac{f(r)}{r^2}\,\left[l(l+n+1) + \frac{n(n+2)}{4} - 
\frac{k(n+2)^2}{4} \left(\frac{r_H}{r}\right)^{n+1}\right]\,,
\label{pot-TV}
\eeq
for tensor- and vector-like perturbations, with $k=-1$ and $k=3$ respectively,
and 
\beq
V_S =  \frac{f(r)}{r^2}\,\frac{q x^3 + p x^2 + w x + z}{4 \left[2m+(n+2)(n+3)x\right]^2}\,,
\label{potential-ten}
\eeq
for scalar gravitational perturbations, with
\beq
m \equiv l(l+n+1) - n -2\,, \qquad
x \equiv \left(\frac{r_H}{r}\right)^{n+1} = 1-f\,,
\eeq
and
\bea
q &\equiv& (n+2)^4 (n+3)^2\,, \qquad 
z \,\,\,\equiv\,\,\, 16 m^3 + 4m^2 (n+2)(n+4)\,, \nonumber\\[2mm]
p &\equiv& (n+2) (n+3) \left[4m\,(2n^2+5n+6) +n (n+2) (n+3) (n-2)\right],\\[2mm]
w &\equiv& -12m\,(n+2)\,\left[m (n-2) + n (n+2) (n+3)\right]. \nonumber
\eea


\section{Solving the Field Equations}

In this section, we proceed to solve analytically the above equation for all
three types of gravitational perturbations in the vicinity of a $(4+n)$-dimensional
neutral, spherically-symmetric black-hole. For this purpose, we will use
a well-known approximate method and solve first the aforementioned equation
at the two asymptotic radial regimes: close to the black hole horizon 
($ r \simeq r_H$), and far away from it ($r \gg r_H$). The two solutions
will then be stretched and matched in an intermediate zone to create a smooth
analytical solution extending over the whole radial regime.

\subsection{The Near-Horizon Regime}

Here, we focus on the derivation of the solution for all types of gravitational
perturbations in the asymptotic regime close to the black hole horizon. Due to
the similar form of their potential, tensor and vector perturbations will be
treated together, while the scalar perturbations will be dealt with separately. 

\bigskip\noindent
{\it A. Tensor and Vector Perturbations}
\medskip

\noindent We start by making the change of variable $r \rightarrow f(r)$,
in terms of which the field equation for tensor and vector perturbations can be
written in the form
\begin{eqnarray}
&~& \hspace*{-0.6cm} 
f\,(1-f)\,\frac{d^2 \Phi}{df^2} + \left[1-\frac{(2n+3)}{(n+1)}\,f \right] \frac{d \Phi}{df}
+ \nonumber \\[2mm]
&~& \hspace*{3cm} \biggl[\,\frac{(\omega r_H)^2}{(n+1)^2 f (1-f)} 
- \frac{A}{(1-f)}+
\frac{k(n+2)^2}{4(n+1)^2}\,\biggr] \Phi=0\,.
\label{NH-1}
\end{eqnarray}
In the above, $k=-1$ for tensor perturbations and $k=3$ for vector ones, and we have
also defined for convenience the quantity
\beq
A \equiv \frac{l(l+n+1)}{(n+1)^2} + \frac{n(n+2)}{4(n+1)^2}\,.
\label{A}
\eeq

If we now make the following field redefinition: $\Phi(f)=f^\alpha (1-f)^\beta F(f)$,
the above equation takes the form of a hypergeometric equation
\beq
f\,(1-f)\,\frac{d^2 F}{df^2} + [c-(1+a+b)\,f]\,\frac{d F}{df} -ab\,F=0\,,
\label{hyper}
\eeq
under the identifications 
\begin{eqnarray}
a=\alpha + \beta + \frac{(n+2)}{2 (n+1)} + G\,, \qquad 
b=\alpha + \beta + \frac{(n+2)}{2 (n+1)} - G\,, \qquad c=1 + 2 \alpha\,,
\label{indices}
\end{eqnarray}
where $G$ an arbitrary constant. Demanding further that the coefficient of
$F(f)$ in our equation be indeed $-ab$ yields three more additional constraints that 
determine the remaining unknown constants, $\alpha$, $\beta$ and $G$.
Their corresponding values are found to be
\beq
G^{(T,\,V)}=\frac{(1+k)(n+2)}{4(n+1)}\,, \qquad \alpha_\pm=\pm \frac{i\omega r_H}{n+1}\,,
\label{alpha-eq}
\eeq
and
\beq
\beta=\frac{1}{2 (n+1)}\,\left\{-1 \pm \sqrt{(2l+n+1)^2-4 \omega^2 r_H^2}\right\}\,.
\label{beta-eq}
\eeq
The indices of the hypergeometric equation $(a,b,c)$ are now fully determined, which
allows us to write the general solution of the hypergeometric equation, and thus the
solution in the near-horizon regime, as
\begin{eqnarray}
&& \hspace*{-1cm}\Phi_{NH}(f)=A_1 f^{\alpha}\,(1-f)^\beta\,F(a,b,c;f)
\nonumber \\[1mm] && \hspace*{2cm} +\,
A_2\,f^{-\alpha}\,(1-f)^\beta\,F(a-c+1,b-c+1,2-c;f)\,,
\label{NH-gen}
\end{eqnarray}
where $A_{1,2}$ are arbitrary integration constants. There is, however, a boundary
condition that this general solution must satisfy: no outgoing waves must be found
right outside the black-hole horizon, as nothing can escape from within this area.
To ensure this, we expand our solution in the limit $r \rightarrow r_H$, or
equivalently  $f \rightarrow 0$, in which case we find
\beq
\Phi_{NH}(f) \simeq A_1\,f^\alpha + A_2\,f^{-\alpha}=
A_1\,e^{-i\omega y} + A_2\,e^{i \omega y}\,,
\label{asym1}
\end{equation}
where, in the last part of the above equation, we have used the `tortoise-like'
coordinate
\beq
\frac{dy}{dr} = \frac{r_H^{n+2}}{f(r)\,r^{n+2}}\,.
\label{tortoise}
\eeq
The asymptotic solution is therefore written in terms of incoming and outgoing
plane waves, as expected, since very close to the horizon the potential $V$ for
all types of gravitational perturbations vanishes. Nevertheless, the
aforementioned boundary condition forces us to discard the outgoing wave by
setting $A_2=0$, that brings the near-horizon solution to its final form
\beq
\Phi_{NH}(f)=A_1 f^{\alpha}\,(1-f)^\beta\,F(a,b,c;f)\,.
\label{NH-fin}
\eeq

The only task remaining is to fix the arbitrary signs appearing in the expressions
of both $\alpha$ and $\beta$. As we may see from Eq. (\ref{asym1}), the interchange
$\alpha_+ \leftrightarrow \alpha_-$ would simply interchange the integration
coefficients $A_1 \leftrightarrow A_2$, therefore the sign of $\alpha$ can be
chosen at random; here, we have chosen $\alpha=\alpha_-$. On the other hand,
the sign in $\beta$ can be fixed by the convergence condition of the hypergeometric
function, i.e. ${\bf Re}\,(c-a-b) = -\frac{1}{n+1} -2 \beta >0$, 
that clearly demands that we choose $\beta=\beta_{-}$.


\bigskip\smallskip\noindent
{\it B. Scalar Perturbations}
\medskip

\noindent By employing the same change of variable, the field
equation for scalar gravitational perturbations can be brought to the form
\begin{eqnarray}
&~& \hspace*{-0.6cm} 
f\,(1-f)\,\frac{d^2 \Phi}{df^2} + \left[1-\frac{(2n+3)}{(n+1)}\,f \right]
\frac{d \Phi}{df} + \nonumber \\[2mm]
&~& \hspace*{3cm} \biggl[\,\frac{(\omega r_H)^2}{(n+1)^2 f (1-f)} 
- \frac{z}{16 (n+1)^2 m^2 (1-f)} - C\,\biggr] \Phi=0\,,
\label{NH-sc}
\end{eqnarray}
where we have defined the quantity
\begin{equation}
C \equiv \frac{q (1-f)^2 + \tilde p (1-f) + \tilde w}{4 (n+1)^2
[2m + (n+2) (n+3) (1-f)]^2}\,, \label{C}
\end{equation}
with
\beq
\tilde p = p -\frac{z (n+2)^2 (n+3)^2}{4 m^2}\,, \qquad 
\tilde w = w - \frac{z (n+2) (n+3)}{m}\,.
\eeq
In this form, Eq. (\ref{NH-sc}) has poles at $f=0$ and $f=1$ (or, at $r=r_H$ and 
$r=\infty$, respectively) while the quantity $C$ takes on a constant value
in both limits.

We then follow a similar method as before, and bring again Eq. (\ref{NH-sc}) to
the form of a hypergeometric equation, with the indices
$(a,b,c)$ given by Eqs. (\ref{indices}), and the powers $(\alpha, \beta)$ by
Eqs. (\ref{alpha-eq})-(\ref{beta-eq}). The only difference arises in the value
of the arbitrary constant $G$ that, in the case of scalar perturbations,
takes the value
\beq
G^{(S)}=\frac{1}{2(n+1)}\,\sqrt{(n+2)^2 - \frac{q + \tilde p + \tilde w}
{[2m + (n+2) (n+3)]^2}}\,.
\label{G-s}
\eeq

By applying the same boundary condition of no-outgoing waves near the horizon,
the general solution for scalar perturbations in the near-horizon regime 
is given again by Eq. (\ref{NH-fin}), with $\beta=\beta_-$ and
$\alpha=\alpha_-$, as before.

\subsection{The Far-Field Regime}

We now turn to the far-field regime. In the limit $r \gg r_H$, $f \rightarrow 1$,
and the field equation for all types of gravitational perturbations -- tensor,
vector and scalar -- takes the simplified form
\beq
\frac{d^2 \Phi}{dr^2} + \left(\omega^2 -\frac{ (n+1)^2 A}{r^2}\right) \Phi=0\,,
\label{FF-eq}
\eeq
where $A$ was defined in Eq. (\ref{A}). By further setting $\Phi=\sqrt{r}\,R$, the new
radial function $R$  is found to satisfy a Bessel differential equation 
\beq
\frac{d^2 R}{dz^2}+ \frac{1}{z}\,\frac{dR}{dz} + \left(1 -
\frac{\nu^2}{z^2}\right)R=0\,,
\label{FF-eq2}
\eeq
with $z=\omega r$ and $\nu = l + (n+1)/2$. Therefore, the general solution of
Eq. (\ref{FF-eq}), standing for the analytical solution in the far-field regime, can
be written as
\beq
\Phi_{FF}(r)=B_1 \sqrt{r}\,J_{l+(n+1)/2}\,(\omega r) + 
B_2 \sqrt{r}\,Y_{l+(n+1)/2}\,(\omega r)\,, \label{FF}
\eeq
where $J_\nu$ and $Y_\nu$ are the Bessel functions of the first and second kind,
respectively, and $B_{1,2}$ arbitrary integration constants.

\subsection{Constructing the complete solution}

Having derived the two asymptotic analytical solutions near the black-hole horizon
and infinity, we may now proceed to construct a solution valid at the whole radial regime
by smoothly connecting the two solutions at an intermediate point. To this end, we
need to stretch the near-horizon solution (\ref{NH-fin}) towards large values of
$r$, and the far-field solution (\ref{FF}) towards small values of $r$. 

Before, however, being able to do the former, we need to rewrite the near-horizon solution
in an alternative form where the argument of the hypergeometric function has been
shifted from $f$ to $1-f$. To this end, we use the standard formula 
\cite{Abramowitz}
\bea
F(a,b,c;f) &=& \frac{\Gamma(c)\,\Gamma(c-a-b)}
{\Gamma(c-a)\,\Gamma(c-b)}\,F(a,b,a+b-c+1;1-f) \\[3mm]
&+& (1-f)^{c-a-b}\,\frac{\Gamma(c)\,\Gamma(a+b-c)}
{\Gamma(a)\,\Gamma(b)}\,F(c-a,c-b,c-a-b+1;1-f) \nonumber 
\eea
in Eq. (\ref{NH-fin}), and subsequently take the limit $r \gg r_H$, or
$f \rightarrow 1$. We then find
\beq
\Phi_{NH}(r) \simeq A_1\,\left(\frac{r_H}{r}\right)^{\beta\,(n+1)}\,
\frac{\Gamma(c)\,\Gamma(c-a-b)}{\Gamma(c-a)\,\Gamma(c-b)} 
+ A_1\,\left(\frac{r_H}{r}\right)^{-1-\beta\,(n+1)}\,\frac{\Gamma(c)\,
\Gamma(a+b-c)}{\Gamma(a)\,\Gamma(b)}\,. \label{stre-NH}
\eeq

We now turn to the FF solution, which we expand in the opposite limit of 
$r \rightarrow 0$. By using standard formulae for the Bessel functions
\cite{Abramowitz}, we then obtain
\beq
\Phi_{FF}(r) \simeq B_1 \left(\frac{\omega}{2}\right)^{l+\frac{n+1}{2}}\,
\frac{r^{l+\frac{n}{2}+1}}{\Gamma\left(l+\frac{n+3}{2}\right)} -
\frac{B_2}{\pi} \left(\frac{2}{\omega}\right)^{l+\frac{n+1}{2}}\,
\frac{\Gamma\left(l+\frac{n+1}{2}\right)}{r^{l+\frac{n}{2}}}\,. \label{stre-FF}
\eeq

Although we have brought both asymptotic solutions in a power-like form,
the powers are not the same. One way to simplify the matching procedure
is to take the low-energy limit $\omega r_H \ll 1$ in the expression of
$\beta$, Eq. (\ref{beta-eq}). At first-order approximation, the $\omega r_H$-term
may be ignored, in which case the powers in Eqs. (\ref{stre-NH}) and 
(\ref{stre-FF}) become identical. A smooth matching is then achieved,
and a complete solution is constructed, if the following relations between
the near-horizon and far-field integration constants hold
\bea
&& \frac{B_1}{A_1}=\left(\frac{2}{\omega r_H}\right)^{l+\frac{n+1}{2}}\,
\frac{\Gamma\left(l+\frac{n+3}{2}\right)\,\Gamma(c)\,\Gamma(c-a-b)}{\Gamma(c-a)\, 
\Gamma(c-b)\,\sqrt{r_H}}\,,
\\[3mm]
&& \frac{B_2}{A_1}=-\pi \left(\frac{\omega r_H}{2}\right)^{l+\frac{n+1}{2}}\,
\frac{\Gamma(c)\,\Gamma(a+b-c)}{\Gamma\left(l+\frac{n+1}{2}\right)\,\Gamma(a)\,
\Gamma(b)\,\sqrt{r_H}}\,.
\eea

The above completes the derivation, in an analytical way, of the solution for
all types of gravitational perturbations in a $(4+n)$-dimensional Schwarzschild
black hole background in the low-energy regime. We now proceed to the calculation
of the corresponding absorption coefficient.

\section{The Absorption Coefficient}

The far-field solution (\ref{FF}) can also be expanded in the limit
$r \rightarrow \infty$, where the effective potential for all types of
gravitational perturbations again vanishes due to the asymptotically flat
behaviour of the metric. In this asymptotic regime, we therefore expect
the general solution to be described again by incoming and outgoing plane
waves. Indeed, in the limit $r \rightarrow \infty$, we obtain
\beq
\Phi_{FF}(r) \simeq \frac{1}{\sqrt{2 \pi \omega}}\,\left\{(B_1-iB_2)\,e^{i \left(\omega r
-\frac{\pi}{2}\,\nu -\frac{\pi}{4}\right)} + (B_1+iB_2)\,e^{-i \left(\omega r
-\frac{\pi}{2}\,\nu -\frac{\pi}{4}\right)}\right\} + ...\,,
\eeq
where as before $\nu=l + (n+1)/2$. From the above expression, we can easily define
the reflection coefficient ${\cal R}_l$ as the ratio of the amplitude of the
outgoing wave over the one of the incoming wave. Then, the absorption probability
follows from the relation
\beq
|{\cal A}_l|^2 = 1- |{\cal R}_l|^2 = 1-\left|\frac{B- i}{B+i}\right|^2\,,
\label{abs}
\eeq
where $B$ is defined as
\beq
B \equiv \frac{B_1}{B_2}=-\left(\frac{2}{\omega r_H}\right)^{2l+n+1}\,
\frac{\Gamma\left(l+\frac{n+3}{2}\right)\,\Gamma\left(l+\frac{n+1}{2}\right)\,
\Gamma(a)\,\Gamma(b)\,\Gamma(c-a-b)\,}{\pi\,\Gamma(c-a)\,\Gamma(c-b)\,\Gamma(a+b-c)}\,.
\label{B}
\eeq
The above two equations comprise our main analytical result for the absorption
coefficient associated with the propagation of gravitons in the higher-dimensional
black hole background given in Eq. (\ref{bhmetric}). Individual solutions for 
scalar, vector and tensor type of perturbations may easily follow upon substituting
the corresponding values for the hypergeometric indices $(a,b,c)$ found in the
previous section. 

\subsection{Simplified analytical result}

The expression (\ref{B}) can be simplified, and a compact analytical result for
the absorption coefficient may thus be derived, if we further expand it in the
low-energy limit $\omega r_H \ll 1$. For convenience, we re-write the 
hypergeometric indices as
\beq
a=\alpha + \beta + G_1\,, \qquad b=\alpha + \beta + G_2\,, \qquad c=1+2 \alpha\,,
\eeq
where
\beq
G_1 \equiv \frac{n+2}{2 (n+1)} + G\,, \qquad G_2 \equiv \frac{n+2}{2 (n+1)} - G\,.
\label{G12}
\eeq
As $G$ takes a different value for scalar, vector and tensor type of gravitational
perturbations, $G_{1,2}$ will also depend on the type of perturbation studied.
We also write Eq. (\ref{abs}) in the form
\beq
|{\cal A}_l|^2 =\frac{2i(B^*-B)}{BB^* +i(B^*-B) +1}\,,\label{abs1}
\eeq
and note that, in the low-energy limit, $BB^* \gg i(B^*-B) \gg 1$. Therefore, by
keeping only the dominant term in the denominator, we arrive at 
\beq
|{\cal A}_l|^2 =K(\omega r_H, \beta)\,\left[Z(\alpha, \beta)- 
Z^*(\alpha, \beta)\right]\,,\label{abs2}
\eeq
where
\beq
K(\omega r_H, \beta) \equiv -\left(\frac{\omega r_H}{2}\right)^{2l+n+1}\,
\frac{2i\pi\,\left(l+\frac{n+1}{2}\right)}{\Gamma\left(l+\frac{n+3}{2}\right)^2}\,
\frac{\Gamma(-1+2\beta+G_1+G_2)}{\Gamma(1-2\beta-G_1-G_2)}
\label{K}
\eeq
and
\beq
Z(\alpha, \beta) \equiv \frac{\Gamma(1+\alpha-\beta-G_1)\,
\Gamma(1+\alpha-\beta-G_2)}{\Gamma(\alpha + \beta + G_1)\,
\Gamma(\alpha + \beta + G_2)}\,.
\label{Z}
\eeq

Let us focus first on the expression for $K(\omega r_H, \beta$).  By using the
following identity satisfied by the Gamma functions \cite{Abramowitz}
\beq 
\Gamma (x)\,\Gamma (1 - x) = -x\,\Gamma (-x)\,\Gamma (x) =  
\frac{\pi}{\sin (\pi x)}\,,
\label{ena} \eeq
keeping the dominant term $\beta^{(0)}$ in the expansion of $\beta$
(Eq. \ref{beta-eq}) in the limit $\omega r_H \ll 1$, i.e.
\beq
\beta^{(0)}  \equiv  - \frac{(2l+n+2)}{2(n + 1)}\,,
\label{beta0}
\eeq
and using ({\ref{G12}}), we find
\beq
K(\omega r_H, \beta) = \left(\frac{\omega r_H}{2}\right)^{2l+n+1}\,
\frac{i\pi^2\,(n+1)}{\Gamma\left(l+\frac{n+3}{2}\right)^2\,
\Gamma\left(1+\frac{2l}{n+1}\right)^2
\,\sin\left[\pi (2 \beta^{(0)} +G_1+G_2)\right]}\,.
\label{K1}
\eeq

The only complex quantity appearing in the expression of $Z(\alpha, \beta)$ is
$\alpha$, which is pu\-re\-ly imaginary, therefore $Z^*(\alpha, \beta)=Z(-\alpha, \beta)$.
By using again Eq. (\ref{ena}), we may write
\bea
&~& \hspace*{-0.4cm}Z-Z^*= \frac{\pi^2}{|\Gamma(\alpha+\beta+G_1)|^2\,
|\Gamma(\alpha+\beta+G_2)|^2}\,\\[3mm]
&~& \hspace*{-0.4cm}\times \,\,\frac{\sin\left[\pi(\alpha+\beta+G_1)\right]\,
\sin\left[\pi(\alpha+\beta+G_2)\right] - \sin\left[\pi(\alpha-\beta-G_1)\right]
\,\sin\left[\pi(\alpha-\beta-G_2)\right]}{|\sin\left[\pi(\alpha+\beta+G_1)\right]|^2
\,\,|\sin\left[\pi(\alpha+\beta+G_2)\right]|^2}\,.\nonumber
\eea
Expanding again in the limit $\omega r_H \ll 1$, or equivalently $\alpha \rightarrow 0$,
we obtain
\beq
Z-Z^* = \frac{2 \alpha}{\pi}\,\sin\bigl[\pi (2 \beta^{(0)} +G_1+G_2)\bigr]\,
\Gamma(1-\beta^{(0)}-G_1)^2\,\Gamma(1-\beta^{(0)}-G_2)^2\,.
\label{Zfinal}
\eeq

By putting Eqs. (\ref{K1}) and (\ref{Zfinal}) together, we finally obtain the
following expression for the absorption probability in the asymptotic low-energy 
regime
\beq
|{\cal A}_l|^2 = 4\pi \left(\frac{\omega r_H}{2}\right)^{2l+n+2}\,
\frac{\Gamma\left(1+\frac{l}{n+1}-G\right)^2\,\Gamma\left(1+\frac{l}{n+1}+G\right)^2}
{\Gamma\left(l+\frac{n+3}{2}\right)^2\,\Gamma\left(1+\frac{2l}{n+1}\right)^2}\,.
\label{abs3}
\eeq
In the above, we have used again the zero-order approximation (\ref{beta0}) for
$\beta$, and the definitions (\ref{G12}) to recover the dependence on the parameter
$G$. According to Eq. (\ref{alpha-eq}), the value of the latter parameter is
zero for tensor gravitational perturbations and $\frac{n+2}{n+1}$ for vector ones,
while for scalar gravitational perturbations its value is given in Eq. (\ref{G-s}). 
In the case of tensor perturbations, it can be shown that the above result
reduces to the one for the absorption probability for a scalar field propagating
in the bulk \cite{kmr1}. This result should have been anticipated by looking at
the equation satisfied by the tensor-like gravitational perturbations in the bulk:
starting from Eq. (\ref{tensor}) and setting $\Phi^{(T)}(r)= \sqrt{r^{n+2}}\,\Phi(r)$,
the new radial function $\Phi(r)$ is found to satisfy the equation 
\beq
\frac{f(r)}{r^{n+2}}\,\frac{d\,}{d r} \left[f(r)\,r^{n+2}\,\frac{d \Phi}{dr}\right]
+\left[\omega^2 -\frac{f(r)}{r^2}\, l(l+n+1) \right] \Phi(r)=0\,,
\eeq
that describes indeed the propagation of a scalar field in the black-hole
background of Eq. (\ref{bhmetric}) \cite{kmr1}.

\begin{table}[t]
\begin{center}
$\begin{array}{ccccc}  \hline \hline
{\rule[-4mm]{0mm}{10mm}
\hspace*{0.4cm} {\bf n} \hspace*{0.6cm}} & \hspace*{0.6cm} {\bf l} \hspace*{0.6cm} &
\hspace*{1.3cm} {\bf |{\cal A}_l^{(T)}|^2} \hspace*{1.2cm} & \hspace*{1.3cm}
{\bf |{\cal A}_l^{(V)}|^2} \hspace*{1.4cm} & \hspace*{1.3cm} {\bf |{\cal A}_l^{(S)}|^2}
\hspace*{0.8cm} \\ \hline
{\rule[-2mm]{0mm}{7mm} n=2} & l=2 & 1.7\cdot 10^{-4}\,(\omega r_H)^8\, &
7.4\cdot 10^{-3}\,(\omega r_H)^8\, & 2.7\cdot 10^{-3}\,(\omega r_H)^8 \\ 
{\rule[-2mm]{0mm}{7mm} } & l=3 & 1.1\cdot 10^{-6}\,(\omega r_H)^{10} &
1.6\cdot 10^{-5}\,(\omega r_H)^{10} & 1.8\cdot 10^{-5}\,(\omega r_H)^{10}\\ 
{\rule[-2mm]{0mm}{7mm} } & l=4 & 4.6\cdot 10^{-9}\,(\omega r_H)^{12} &
3.7\cdot 10^{-8}\,(\omega r_H)^{12} & 5.5\cdot 10^{-8}\,(\omega r_H)^{12}
\\ \hline
{\rule[-2mm]{0mm}{7mm} n=4} & l=2 & 3.2\cdot 10^{-6}\,(\omega r_H)^{10} &
2.2 \cdot 10^{-4}\,(\omega r_H)^{10} & 3.4\cdot 10^{-5}\,(\omega r_H)^{10} \\ 
{\rule[-2mm]{0mm}{7mm} } & l=3 & 1.9\cdot 10^{-8}\,(\omega r_H)^{12} &
4.2\cdot 10^{-7}\,(\omega r_H)^{12} & 3.1\cdot 10^{-7}\,(\omega r_H)^{12} \\ 
{\rule[-2mm]{0mm}{7mm} } & l=4 & 8.1 \cdot 10^{-11}\,(\omega r_H)^{14} &
\,9.5\cdot 10^{-10}\,(\omega r_H)^{14} & \,1.3\cdot 10^{-9}\,(\omega r_H)^{14}\\ \hline
{\rule[-2mm]{0mm}{7mm} n=6} & l=2 & 3.1\cdot 10^{-8}\,(\omega r_H)^{12}\, &
3.2\cdot 10^{-6}\,(\omega r_H)^{12} & 2.2\cdot 10^{-7}\,(\omega r_H)^{12}\\ 
{\rule[-2mm]{0mm}{7mm} } & l=3 & 1.5\cdot 10^{-10}\,(\omega r_H)^{14} &
4.7\cdot 10^{-9}\,(\omega r_H)^{14} & 1.9\cdot 10^{-9}\,(\omega r_H)^{14} \\
{\rule[-2mm]{0mm}{7mm} } & l=4 & 5.4 \cdot 10^{-13}\,(\omega r_H)^{16} &
\,8.9\cdot 10^{-12}\,(\omega r_H)^{16} & 8.4\cdot 10^{-12}\,(\omega r_H)^{16}
\\ \hline \hline
\end{array}$
\end{center}
\caption{Dependence of the absorption probability for tensor, vector and
scalar gravitational perturbations in the bulk on $n$ and $l$, in the asymptotic
regime $\omega r_H \rightarrow 0$.}
\end{table} 

 From Eq. (\ref{abs3}), one may see that the absorption probability depends on
both the angular momentum number $l$ and number of extra dimensions $n$, through the
arguments of the Gamma functions as well as the power of $\omega r_H$. As either
$l$ or $n$ increases, the latter increases too, which, for $\omega r_H \ll 1$,
causes a suppression in the value of $|{\cal A}_l|^2$. As the behaviour of the
remaining factor is not equally clear, in Table 1 we display the explicit value
of $|{\cal A}_l|^2$ for all three types of gravitational perturbations, as these
follow from the simplified expression (\ref{abs3}), for the indicative
values $n=2,4,6$ and $l=2,3,4$. From these entries, one may easily conclude that,
as either $l$ or $n$ increases, the value of the absorption probability for all
types of gravitational perturbations in the asymptotic low-energy regime is
significantly suppressed.

As a final comment, let us note here that, for the same values of $l$ and $n$,
$|{\cal A}_l|^2$ assumes a different value for different types of gravitational
perturbations. From the entries of Table 1, we may easily see that the tensor
perturbations are by orders of magnitude suppressed compared to the vector and
scalar perturbations, while the relative magnitude of the latter two is strongly
dependent on the particular values of $l$ and $n$. As the tensor perturbations
have the same absorption probability as the bulk scalar fields, it would be
interesting to see whether gravitons dominate over scalar fields during
the emission of Hawking radiation in the bulk. 

\subsection{Complete analytical result}

In deriving our main result for the absorption probability for gravitational
perturbations in the bulk, Eqs. (\ref{abs})-(\ref{B}), the use of the
low-energy assumption was made only once -- during the matching of the two
asymptotic solutions in the intermediate zone. Nevertheless, that was enough
to restrict the validity of our solution to values of the $\omega r_H$
parameter well below unity. The simplified analytical result (\ref{abs3}) was,
on the other hand, the result of a series of expansions in the arguments of
the Gamma functions appearing in Eq. (\ref{B}), and, as a result, its validity
is significantly more restricted. In Table 2, we display the values
of the absorption coefficient derived by using our two analytical expressions,
Eqs. (\ref{abs})-(\ref{B}) and (\ref{abs3}), as $\omega r_H$ ranges from 0.001
to 0.5. We may easily see that, for very low values of $\omega r_H$, the agreement
between the two values is remarkable; however, as soon as $\omega r_H$
reaches the value 0.5, the deviation between the two values appearing in
the last row of Table 2 reaches the magnitude of 15\%. 

\begin{table}[t]
\begin{center}
$\begin{array}{ccc}  \hline \hline
{\rule[-3mm]{0mm}{8mm}
\hspace*{0.5cm} \omega r_H \hspace*{0.5cm}} & \hspace*{0.2cm}
|{\cal A}_l^{(T)}|^2\,\,(\rm simplified\,\,expression)
\hspace*{0.2cm} & \hspace*{0.2cm} |{\cal A}_l^{(T)}|^2\,\,(\rm exact\,\,expression)
\hspace*{0.2cm} \\ \hline
{\rule[-2mm]{0mm}{6mm} 0.001} & 1.6997 \times 10^{-28} & 1.6997 \times 10^{-28} \\ 
{\rule[-2mm]{0mm}{6mm} 0.01} & 1.6997 \times 10^{-20} & 1.6999 \times 10^{-20}\\ 
{\rule[-2mm]{0mm}{6mm} 0.1} & 1.6697 \times 10^{-12} & 1.7112 \times 10^{-12} \\ 
{\rule[-2mm]{0mm}{6mm} 0.3} & 1.1152 \times 10^{-8} & 1.1839 \times 10^{-8} \\ 
{\rule[-2mm]{0mm}{6mm} 0.5} & 6.6396 \times 10^{-7} & 7.8001 \times 10^{-7}\\ \hline \hline
\end{array}$
\end{center}
\caption{Deviation between the values of the absorption probability given by the 
simplified and complete analytical expression, for tensor gravitational
perturbations, for $n=2$, $l=2$ and different values of $\omega r_H < 1$.}
\end{table} 

In this section, we return to our main analytical result for the absorption
probability of gravitons in the bulk, Eqs. (\ref{abs})-(\ref{B}). As it
was shown in the case of emission on the brane \cite{Kanti, HK1},
the analytical, non-simplified, expression for $|{\cal A}_l|^2$ is in
excellent agreement with the exact numerical result in the low-energy
regime and in a good -- both quantitative and qualitative -- agreement
in the intermediate-energy regime. In the high-energy regime, the validity
of our expression naturally breaks down. There, we expect the greybody
factor (absorption cross-section) to be given in terms of the absorptive area
of the black hole: this result has been computed in \cite{HK1, Kanti},
and holds for all types of perturbations independently of their spin.

\begin{figure}[t]
\begin{center}
\thicklines
\mbox{
\includegraphics[height=5.6cm,clip]{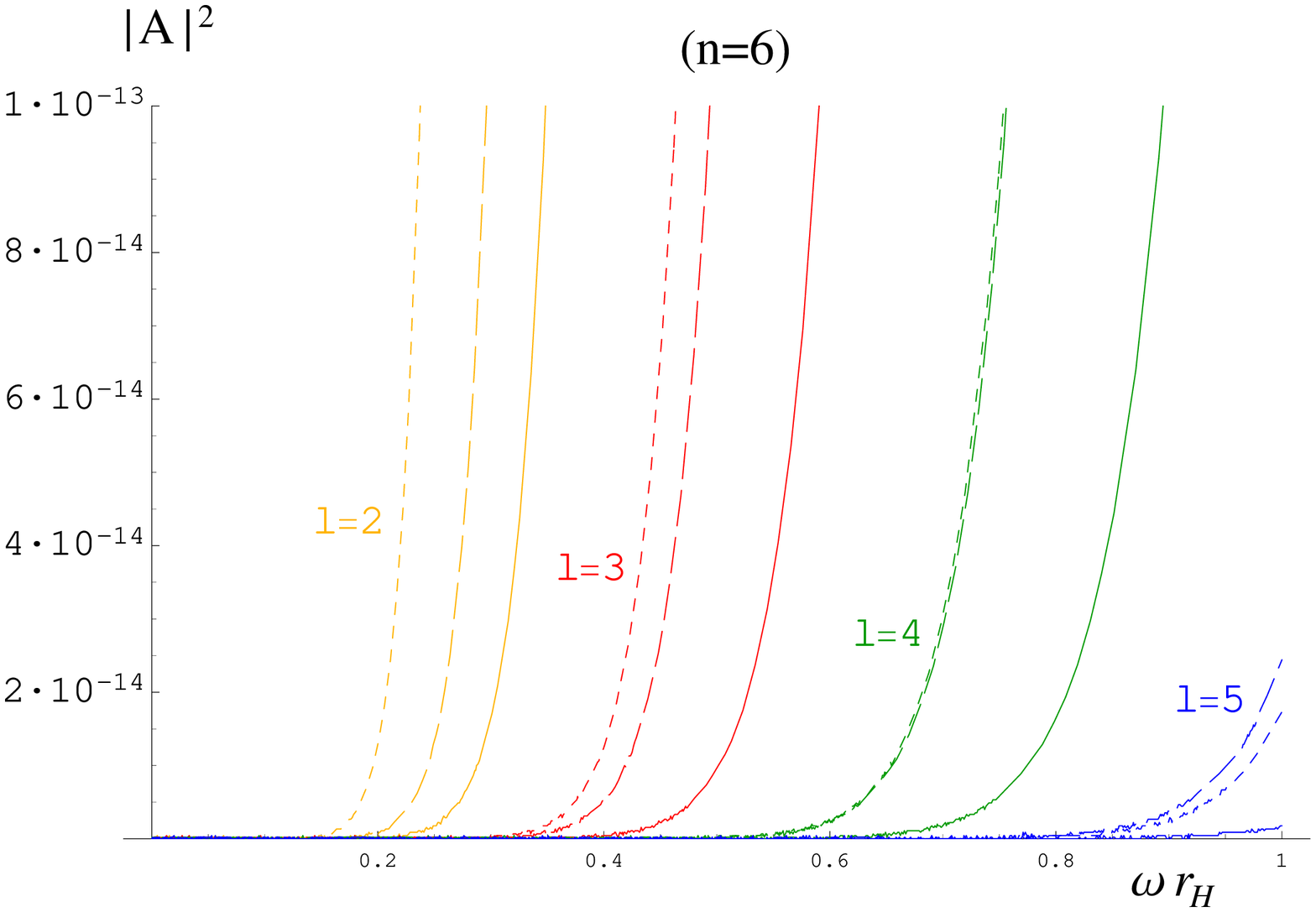}}\hspace*{-0.1cm}
{
\includegraphics[height=5.6cm,clip]{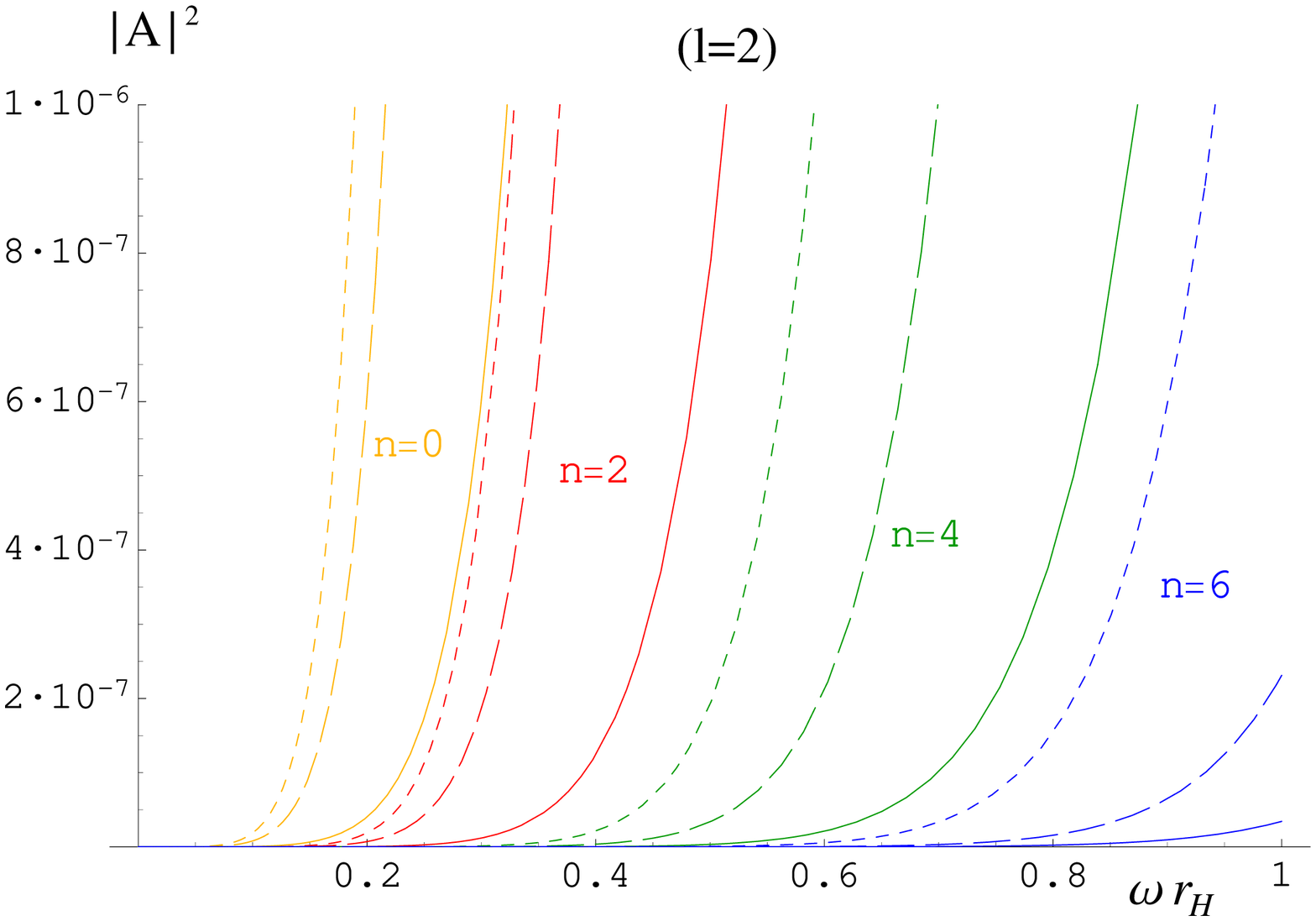}}
\caption{Absorption probability $|{\cal A}_l|^2$ for tensor (solid lines), vector
(short-dashed lines) and scalar (long-dashed lines) gravitational perturbations
in the bulk for: {\bf (a)} $n=6$, and $l=2,3,4,5$, and {\bf (b)} $l=2$, and
$n=0,2,4,6$.}
\label{gravitons-nl}
\end{center}
\end{figure}

In Figs. 1(a) and (b), we depict the absorption probability for all types of
gravitational perturbations, as a function of the dimensionless energy parameter
$\omega r_H$, and for different values of the angular momentum number $l$
and number of extra dimensions $n$, respectively. As expected, for all types
of perturbations and values of $l$ and $n$, the absorption coefficient vanishes
when the energy of the propagating particle goes to zero, while it increases
with $\omega r_H$.  This is in agreement with the classical scattering theory:
the larger its energy, the more likely it is for a particle to overcome the
gravitational field of the black hole and escape to infinity. Figure 1(a) 
reveals that, as $l$ increases, the absorption probability for all types of
gravitational perturbations is suppressed, in accordance with the behaviour
noticed in the entries of Table 1; although the results depicted correspond
to the case $n=6$, this behaviour was found to hold for all values of $n$.
For fixed $l$, on the other hand, as we may see from Fig. 1(b),
$|{\cal A}_l|^2$ is found to exhibit a significant suppression also with $n$, for
all types of gravitational perturbations, in agreement again with the behaviour
noticed in the entries of Table 1. This behaviour is observed for the lowest
two partial modes with $l=2$ and $l=3$; for higher values of $l$,
$|{\cal A}_l|^2$ shows a temporary enhancement, as $n$ increases from 0 to 2,
that nevertheless changes again to a rapid suppression as $n$ increases further.

As it was noticed by studying the asymptotic low-energy regime, the absorption
probability for tensor gravitational perturbations is significantly suppressed,
compared to the other two types, also in the extended low-  and
intermediate-energy regime. 
On the other hand, the vector and scalar perturbations may dominate one over
the other for different values of $l$: as it can be seen from our
figures, the vector-like gravitational perturbations predominantly dominate
for low values of $l$, while the scalar-like perturbations usually take over for
large values of $l$. As the higher partial waves are significantly suppressed,
we are led to conclude that the vector-type perturbations will be the
dominant gravitational degree of freedom to be emitted in the bulk by a
Schwarzschild-like higher-dimensional black hole. However,
different types of perturbations are characterized by a different multiplicity
of states for the same value of $l$, and the latter must be taken into account
before such conclusions can be safely drawn.

\section{Energy Emission Rate}

We finally turn to the energy emission rate by the higher-dimensional
Schwarzschild black hole for gravitons in the bulk. Let us first address the
issue of the multiplicities of states that correspond to the same angular
momentum number $l$. For tensor, vector, and scalar type of perturbations
these are given by the expressions \cite{Rubin}
\beq
N_l^{(T)}=\frac{n(n+3)(l+n+2)(l-1)(2l+n+1)(l+n-1)!}{2 (l+1)!\,(n+1)!}\,,
\label{mult-ten}
\eeq
\smallskip
\beq
N_l^{(V)}=\frac{l(l+n+1)(2l+n+1)(l+n-1)!}{(l+1)!\,n!}\,,
\label{mult-vec}
\eeq
\smallskip
\beq
N_l^{(S)}=\frac{(2l+n+1)(l+n)!}{l!\,(n+1)!}\,,
\label{mult-sc }
\eeq
respectively. As it is evident, the multiplicities depend also on the dimensionality
of spacetime. In Table 3, we display the multiplicities of states for tensor,
vector and scalar perturbations, for some indicative values of $l$ and $n$. We
immediately observe the proliferation of states as either one of these two parameters 
increases, especially the one with the number of extra dimensions. It thus becomes
clear that the value of the absorption probability alone will not be the only
decisive factor that will determine the contribution of each type of gravitational
perturbation to the total emission rate of the black hole. 

\begin{table}[t]
\begin{center}
$\begin{array}{cccc}  \hline \hline
{\rule[-2mm]{0mm}{8mm}
\hspace*{0.5cm}  \hspace*{0.5cm}} & \hspace*{1.3cm} N_l^{(T)} \hspace*{1.3cm} 
& \hspace*{1.3cm} N_l^{(V)} \hspace*{1.3cm} &  \hspace*{1.3cm} N_l^{(S)}
\hspace*{1.3cm} \\ 
{\rule[-3mm]{0mm}{8mm}
\hspace*{1cm}  \hspace*{1cm}} & n=1\,\,\,\,\,\,\,n=6\,\,\,\, & 
n=1\,\,\,\,\,\,\,n=6\,\,\,\, & 
\hspace*{0.2cm} n=1\,\,\,\,\,\,\,n=6\,\,\,\, \hspace*{0.2cm} \\ \hline
{\rule[-2mm]{0mm}{7mm} l=2} & 10\,\, \qquad 495\,\, & 16 \qquad 231 & 9\, \qquad \,44\\
{\rule[-2mm]{0mm}{7mm} l=3} & 24 \qquad 2574 & 30 \qquad 910 & 16 \qquad 156\\ 
{\rule[-2mm]{0mm}{7mm} l=4} & 42 \qquad 8748 & \,48\,\,\, \quad \,2772 & 25 \qquad 450\\ 
{\rule[-2mm]{0mm}{7mm} l=5} & \,64\, \quad \,\,\,23868 & \,70\,\,\, \quad \,7140
& \,36\,\, \quad \,\,\,1122
\\ \hline \hline
\end{array}$
\end{center}
\caption{Multiplicities of states corresponding to the same angular momentum
number $l$ for tensor, vector and scalar perturbations, for $n=1$ and $n=6$.}
\end{table} 

With the multiplicities of states and values of the absorption probability for each
type of perturbation at our disposal, we can now calculate the corresponding energy
emis\-sion rate at the low-energy regime. The contribution of each type of gravitational
perturbation to the total graviton energy emission rate is given by the
expression\cite{Kanti}
\beq
\frac{d^2 E^{(P)}}{dt\,d\omega}=\frac{1}{2\pi}\,\sum_l\,N_l^{(P)}
|{\cal A}_l^{(P)}|^2\,\frac{\omega}{\exp(\omega/T_H) - 1}\,,
\label{hawking}
\eeq
where the superscript $P=(T,V,S)$ denotes the type of perturbation, and the black
hole temperature is given in Eq. (\ref{TH}). Summing over the three contributions,
we may obtain the total amount of energy emitted per unit time and unit frequency
by the black hole in the form of gravitons in the bulk. As this result would be
more useful in the context of an exact numerical analysis (to
which we hope to return soon), here we will concentrate on the relative emission
rates for the different types of gravitational perturbations and their relative
magnitude to the one for bulk scalar fields. 

A simple numerical analysis, by combining Eq. (\ref{hawking}) with the entries
of Tables 1 and 3, reveals that, in the asymptotic low-energy regime, the
vector-like perturbations are indeed the dominant type of gravitational degree
of freedom emitted in the bulk by the black hole. For example, for $n=2$ and
$l=2$, vector perturbations amount to 85\% of the total gravitational degrees
of freedom emitted, compared to 13\% for scalar and 2\% for tensor degrees.
As $n$ increases further, so does the dominance of the vector-like perturbations
that, for $n=6$ and $l=2$, reaches the magnitude of 97\%. This dominance is
never over-turned but it may be significantly decreased, at the level of higher
partial waves: for instance, for $n=6$ and $l=4$, the vector, scalar and tensor
perturbations correspond to 74\%, 12\% and 14\%, respectively, of the total
number of gravitational degrees of freedom emitted. In fact, due to their
large multiplicity of states, the tensor perturbations dominate over the scalar
ones, for most large values of $n$ and/or $l$. 

\begin{figure}[t]
\begin{center}
\thicklines
\mbox{
\includegraphics[height=5.4cm,clip]{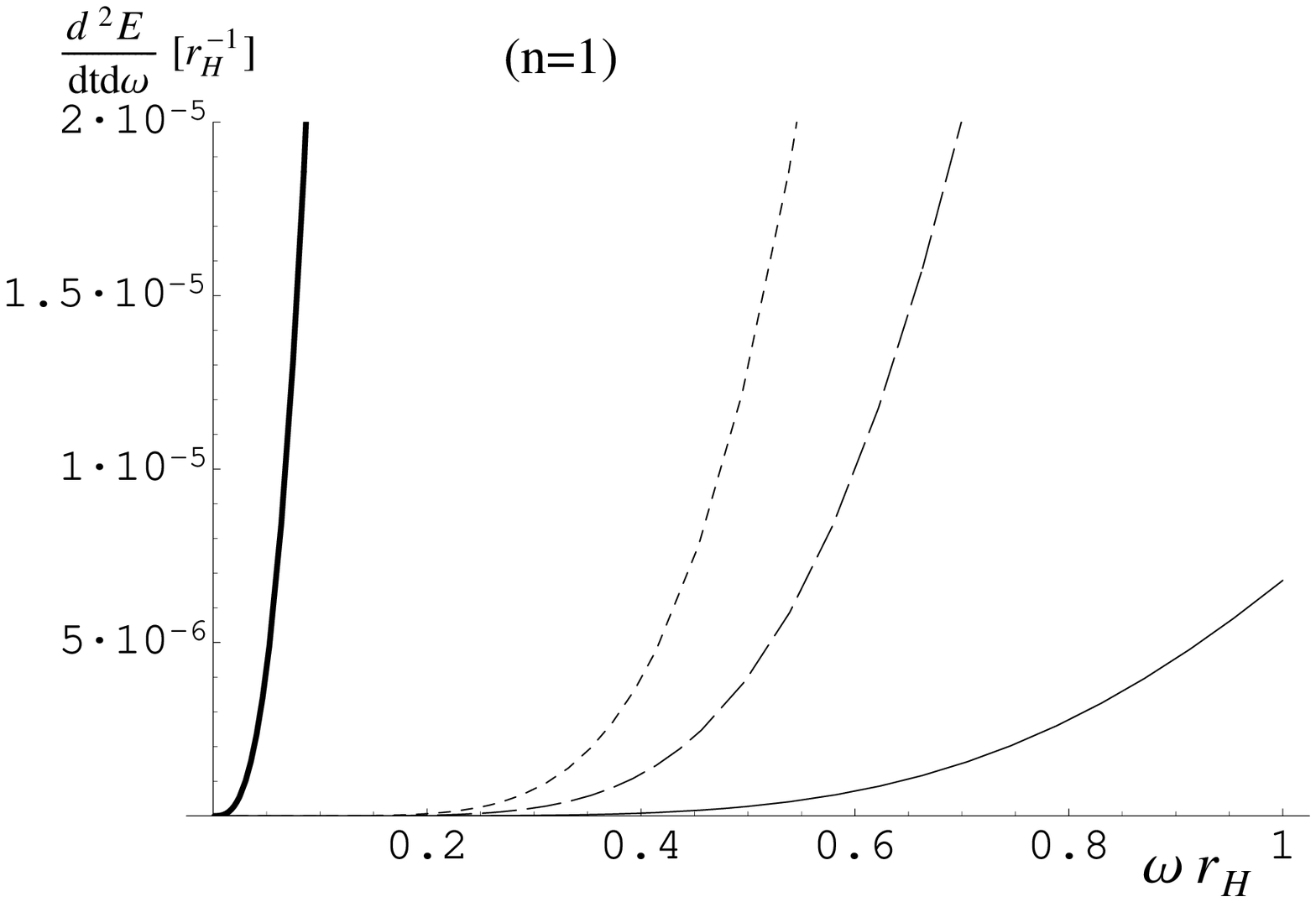}}\hspace*{-0.2cm}
{
\includegraphics[height=5.4cm,clip]{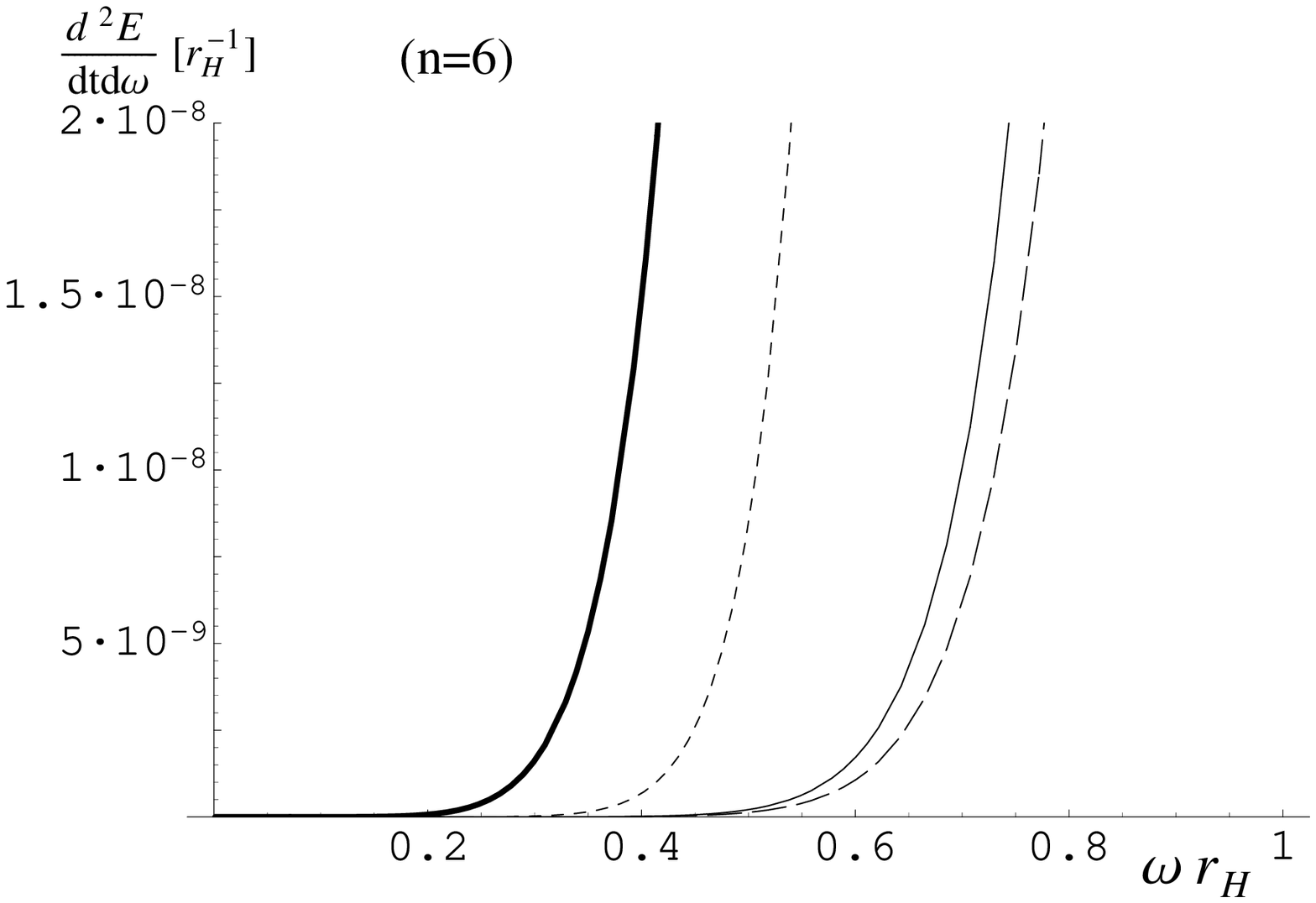}}
\caption{Energy emission rates for tensor (thin solid lines), vector
(short-dashed lines) and scalar (long-dashed lines) gravitational perturbations,
and scalar fields (thick solid lines) in the bulk, for {\bf (a)} $n=1$, and
{\bf (b)} $n=6$.}
\label{rates}
\end{center}
\end{figure}

The above results valid at the lowest part of the radiation spectrum may change
for higher values of the energy parameter. Therefore, in Figs. 2(a) and (b),
we plot the energy emission rates for different types of gravitational
perturbations, for $n=1$ and $n=6$, respectively, as these follow by using
our complete analytical expression (\ref{abs})-(\ref{B}) for the
absorption probability. For comparison, we also display the energy emission
rate for scalar fields in the bulk. In the sum over $l$ in Eq. (\ref{hawking})
we have included all modes up to $l=12$, although at the very low-energy part
of the spectrum, the contribution of all modes with $l \geq 4$ is at least 
four orders of magnitude smaller than the one of the $l=2$ for every value of
$n$. Having included a large enough number of modes
in the sum, allows for deviations from this behaviour
as the energy increases.

The results depicted in Fig. 2 are in fact in excellent agreement with the
conclusions derived by using the simplified expression for $|{\cal A}_l|^2$.
The vector-type perturbations are indeed the dominant gravitational degrees
of freedom to be emitted by the black hole in the bulk for every value of $n$.
What depends on the number of extra dimensions is the relative magnitude of
the energy emission rate of scalar and tensor perturbations: whereas for low $n$,
the tensor perturbations are subdominant to the scalar ones over the entire
low-energy regime, for high $n$, they clearly dominate over the latter ones.
Despite the above, the energy emission rates for all types of gravitational
perturbations in the bulk, even when combined, remain well below the one
for scalar fields in the low-energy regime. This is due to the fact that
the emission rate for scalar fields receives a significant enhancement, in
this particular energy regime, from the dominant $l=0$ and $l=1$ modes, that
are absent from the spectrum of gravitational perturbations. As the energy
parameter increases though, we expect the higher partial waves to gradually
come into dominance and possibly help gravitons to dominate over the
bulk scalar fields.

By comparing finally the vertical axes of the two plots in Fig. 2, we 
conclude that the low-energy emission rate for all types of degrees of freedom
in the bulk decreases as the number of extra dimensions increases. 
Although both the temperature of the black hole and multiplicity of states
undergo a significant enhancement as $n$ increases, the equally significant
suppression of the absorption probability, depicted in Fig. 1(b), prevails,
leading to the suppression of the number of degrees of freedom emitted by
the black hole. This low-energy suppression for bulk scalar fields was
first seen in \cite{kmr1, HK1} but the exact numerical analysis performed
in the latter work showed that, for higher values of the energy parameter,
the spectrum is enhanced with the number of extra dimensions. This is caused
by the milder suppression of the absorption probability with $n$ at higher
frequencies, but also by the shift of the emission curve towards higher
energies, in accordance with Wien's law -- the latter leads to the emission
of more higher frequency particles and less low-energy ones as $n$ increases
\cite{HK1}. Due to the similarities observed in the behaviour of gravitational
and scalar fields in the bulk, we expect the same enhancement to take place
also for gravitons at higher energy regimes.


\section{Conclusions}

A higher-dimensional black hole created in the context of a brane-world
theory will decay through the emission of Hawking radiation both in the bulk and
on the brane. The type of particles emitted along each ``channel" are 
determined by the assumptions of the particular model, with only gravitons and 
possibly scalars propagating in the bulk within the framework of the Large Extra
 Dimensions scenario. Although the emission of scalars has been studied in 
detail, until recently gravitons have received little attention. 

In this work, we have addressed this gap in the literature by investigating the
emission of tensor, vector and scalar gravitational modes in the bulk from 
a ($4+n$)-dimensional Schwarzschild black hole. Working in the low energy regime, 
we analytically solved the corresponding field equations and computed the 
absorption probability in each case. Both a complete analytic expression and its
asymptotic low-energy simplification were studied in detail, and their dependence 
on the angular momentum number $l$ and number of extra dimensions $n$ was
examined. Although numerically different as the energy increases, these two
expressions have identical qualitative behaviours, that reveal an increase
in the value of the absorption probability with increasing energy and
suppression as the number of partial wave $l$ and extra dimensions $n$ increase. 

The complete analytical expression for the absorption probability was then
used to derive the contribution of each gravitational degree of freedom
to the total graviton emission rate of the black hole in the bulk. Accounting
for the rapid proliferation of state multiplicities with $l$ or $n$, a sum
over the first twelve partial waves was performed to obtain the total
low-energy emission rate for each gravitational degree of freedom. Our
results show that vector perturbations are the dominant mode emitted in
the bulk for all values of $n$. The relative emission rates of the subdominant
scalar and tensor modes depend on $n$, with scalars foremost at small $n$ and 
tensors more prevalent at high $n$. The absence of the $l=0,1$ partial waves,
dominant in the low-energy regime, from all gravitational spectra causes even
the total gravitational emission rate to be subdominant to that of scalar fields
in the bulk. Finally, as previously found for bulk scalar fields, the energy
emission rates for all types of gravitational perturbations are suppressed
with the number of extra dimensions in the entire low-energy regime.

In this work we have focused on the low-energy part of the Hawking radiation
spectrum as it permitted analytical calculation and derivation of closed form expressions for the absorption probability for gravitons. Although the radiation 
emitted in the bulk is not directly observable, it determines the energy left
for emission on our brane. In this context, our results, in addition to their 
theoretical interest, would be of particular use to experiments developed to
detect the low-energy spectrum of radiation emitted from a higher-dimensional
black hole. As the exact form of the complete gravitational spectrum is still
pending, we hope to return in the near future with results from a numerical
analysis that could only provide the answer to this question.

{\bf Note added in proof.} While this work was at its last stages, two
relevant papers appeared in the literature, \cite{Park} and \cite{Cardoso}.
In particular, the latter work \cite{Cardoso}, that studies the graviton
emission rate in the bulk, overlaps with ours with respect to the analytical
results for the cases of tensor and vector gravitational perturbations,
which are in agreement with ours. 



\bigskip

{\bf Acknowledgments.} S.C and O.E. acknowledge PPARC and I.K.Y. fellowships,
respectively. The work of P.K. is funded by the UK PPARC Research
Grant PPA/A/S/ 2002/00350. This research was co-funded by the European Union
in the framework of the Program $\Pi Y\Theta A\Gamma O PA\Sigma-II$ 
of the {\textit{``Operational Program for Education and Initial
Vocational Training"}} ($E\Pi EAEK$) of the 3rd Community Support Framework
of the Hellenic Ministry of Education, funded by $25\%$ from 
national sources and by $75\%$ from the European Social Fund (ESF).


\end{document}